# Crystal structure prediction at finite temperatures

*Ivan A. Kruglov* [1,2,*], *Alexey V. Yanilkin* [1,2], *Yana Propad* [1], *and Artem R. Oganov* [3*]

[1] Moscow Institute of Physics and Technology, 9 Institutsky lane, Dolgoprudny 141700, Russia
[2] Dukhov Research Institute of Automatics (VNIIA), Moscow 127055, Russia
[3] Skolkovo Institute of Science and Technology, Skolkovo Innovation Center, 3 Nobel Street, Moscow 121205, Russia

**Corresponding Authors**
*Ivan A. Kruglov, E-mail: ivan.kruglov@phystech.edu
*Artem. R. Oganov, E-mail: a.oganov@skoltech.ru

**Crystal structure prediction is a central problem of theoretical crystallography and materials science, which until mid-2000s was considered intractable. Several methods, based on either energy landscape exploration[1,2] or, more commonly, global optimization[3–8], largely solved this problem and enabled fully non-empirical computational materials discovery[9,10]. A major shortcoming is that, to avoid expensive calculations of the entropy, crystal structure prediction was done at zero Kelvin and searched for the global minimum of the enthalpy, rather than free energy. As a consequence, high-temperature phases (especially those which are not quenchable to zero temperature) could be missed. Here we develop an accurate and affordable solution, enabling crystal structure prediction at finite temperatures. Structure relaxation and fully anharmonic free energy calculations are done by molecular dynamics with a force field (which can be anything from a parametric force field for simpler cases to a trained on-the-fly machine learning interatomic potential), the errors of which are corrected using thermodynamic perturbation theory to yield accurate *ab initio* results. We test the accuracy of this method on metals (probing the *P-T* phase diagram of Al and Fe), a refractory intermetallide (WB), and a significantly ionic ceramic compound (Earth-forming silicate $MgSiO_3$ at pressures and temperatures of the Earth's lower mantle). We find that the *hcp*-phase of aluminum has a wider stability field than previously thought, and the temperature-induced α-β transition in WB occurs at 2789 K. It is also found that iron has *hcp* structure at conditions of the Earth's inner core, and the much debated (and important for constraining Earth's thermal structure) Clapeyron slope of the post-perovskite phase transition in $MgSiO_3$ is 5.88 MPa/K.**

The search for stable crystal structure is a very challenging task, amounting to finding the structure with the lowest free energy among an astronomical number of possible structures, for each of which the free energy has to be computed. Today, this can only be done at zero Kelvin[2–5,7]: relaxing crystal structures and computing their free energies at finite temperatures would require extensive statistical sampling, increasing the number of configurations by several orders of magnitude. This challenge must be addressed, because high-temperature phases often possess interesting properties and are of fundamental and practical interest. Such elements as titanium, iron, calcium, as well as many compounds (for example, cubic paraelectric $BaTiO_3$ or ultrahard borides, carbides and nitrides) have high-temperature phases which have no stability fields at zero temperature and some may not even be metastably quenched at low temperatures. Another important example is planetary science: properties of planet-forming materials at extreme conditions of planetary interiors are responsible for the observed seismic profiles and dynamical processes occurring within planets. For example, the exact structure of iron in the Earth's core[11–14] and exact location of the perovskite - post-perovskite $MgSiO_3$ boundary at conditions of the lower mantle[15–19] are still under debate.

Here we develop a new method, T-USPEX, bringing crystal structure prediction beyond the zero Kelvin regime. It is based on a previously developed evolutionary crystal structure prediction method USPEX[3–5], which predicts lowest-energy crystal structure for a given chemical composition. Below we describe the premises on which we build our approach.

First, while the need for sampling increases computing costs by several, roughly 3-5, orders of magnitude, there are reasons to hope that this can be overcome. As temperature increases, the number of local minima of the free energy decreases, as minima separated by low barriers merge (Fig. 1) and the free energy landscape becomes simpler and smoother. This decrease must be very fast (at zero temperature the number of minima $C_0$ is very large, let's say, $>>10^6$ for medium-size systems, whereas near the melting temperature there is only one or very few local minima). This has led Oganov[20] to conjecture exponential decrease of the number of local minima $C$:

$$C = C_0 \exp\left(\alpha \frac{T^* - T}{T^*}\right), \qquad (1)$$

where $\alpha$ is a constant, and $T^*$ is a characteristic temperature (higher than the melting temperature), at which only one free energy minimum exists. Such rapid decrease of the number of minima makes the problem much simpler: the number of structure relaxations and free energy calculations needed to find the global minimum decreases strongly.

Second, each structure relaxation and free energy calculation can be made much cheaper if one replaces *ab initio* calculations with a machine learning (ML) force field[21–25], which speeds up the calculations by at least two orders of magnitude[26] with only a small reduction of accuracy (usual mean average errors are ~10 meV/atom[26]). Moreover, with ML one can consider large supercells, thus taking into account long-wavelength phonons, which are so important for thermodynamics. ML force fields have been applied to many different problems[21–25,27–32], for a comparison of different ML approaches, see Ref.[33]. ML force fields were used to build phase diagram of uranium[32], accelerate crystal structure prediction method[26], calculate thermal conductivity[34], vibrational properties[35] and so on. Here we mostly use the MTP potential[36], which showed outstanding performance[33]. Structure relaxation at given *P-T* conditions is equivalent to taking statistical averages for lattice translation vectors and for atomic fractional coordinates, while Gibbs free energy is calculated for each structure as:

$$G = PV + F = PV + F_0 + \int_0^1 (U(\lambda) - U_0) d\lambda, \qquad (2)$$

where $P$ is the external pressure, $V$ the equilibrium volume, and $F$ is the Helmholtz free energy which is computed by thermodynamic integration using adiabatic switching from a reference system with known free energy $F_0$ and potential $U_0$ (we use the Einstein crystal as reference system) to the system with potential $U(\lambda=1)$ whose free energy we wish to compute[37,38].

It is very important that we correct the errors of the ML potential to obtain free energies equivalent to full *ab initio* values $F_{AI}$ to second order using thermodynamic perturbation theory[39]:

$$F_{AI} \simeq F + \frac{1}{N_{at}} \left[ \langle U_{AI} - U \rangle - \frac{1}{2k_b T} \langle [U_{AI} - U]^2 \rangle \right], \qquad (3)$$

where $U_{AI}$ is the total energy of the ab initio system, $k_B$ and $T$ are Boltzmann constant and temperature, respectively, and "$\langle \ \rangle$" denotes ensemble averages. Free energy calculations (2) are done on large supercells with ~1000 atoms, in the NVT ensemble (typical duration of a run is 40 ps). Free energy corrections (3) are done on smaller ~60-atom supercells (this required a classical NVT run of 10 ps duration, where DFT energies and ensemble averages in (3) were computed for 100 equally spaced snapshots).

Here we take advantage of both these factors – the reduction of complexity of the free energy surface on increasing temperature, and the possibility of computing accurate free energies in an economic way via ML force field using thermodynamic integration and thermodynamic perturbation theory corrections. This allowed us to develop an efficient and reliable approach for crystal structure prediction at finite temperatures, and we illustrate its performance on four substances – aluminum (calculating its phase diagram and searching for possibly missed unknown phases), iron (to figure out the stable crystal structure at conditions of the Earth's inner core), $MgSiO_3$ (calculating its important phase diagram and searching for possibly overlooked phases at conditions of the lowermost mantle of the Earth), and WB (verifying T-USPEX method for a complex multicomponent system and predicting stability of a temperature-induced phase).

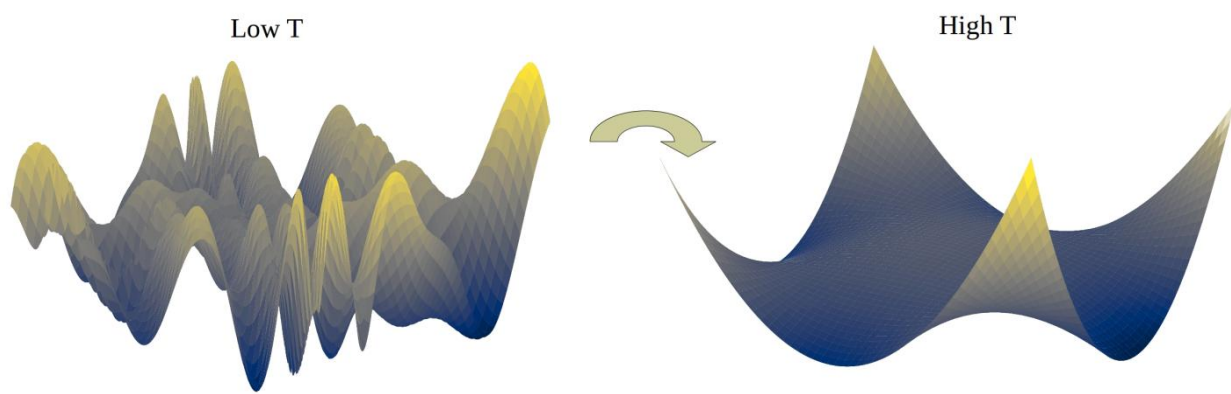

Fig. 1. Schematic representation of the free energy surface at low and high temperature.

For aluminum, three phases are experimentally known to be stable at different P-T conditions: *fcc* (at normal conditions), *hcp* and *bcc*. First, using DFT we determined that at zero Kelvin fcc-Al transforms to hcp-Al at 162 GPa (Fig. S1), and hcp-Al transforms to bcc-Al at 373 GPa (Fig. S2). These values agree well with previous theoretical calculations[40–42], but in experiment fcc-Al transforms to hcp-Al at 217 ± 10 GPa[43] at 300 K. Then, we ran T-USPEX calculations for Al at 0 GPa and 300 K, 100 GPa and 2000 K, 400 GPa and 2000 K. Each generation consisted of 20 randomly generated structures, and used the MTP potential trained on results of *ab initio* molecular dynamics (see **Methods** section). We saw that during subsequent *NPT* molecular dynamics calculations (used for relaxing structures at finite temperatures) almost all randomly generated structures transformed either to the fcc, hcp and/or bcc structures, or to defective structures with large unit cells (the latter structures were discarded). The transformation of a randomly generated Al structure to bcc-Al in a T-USPEX calculation is shown in Fig. 2.

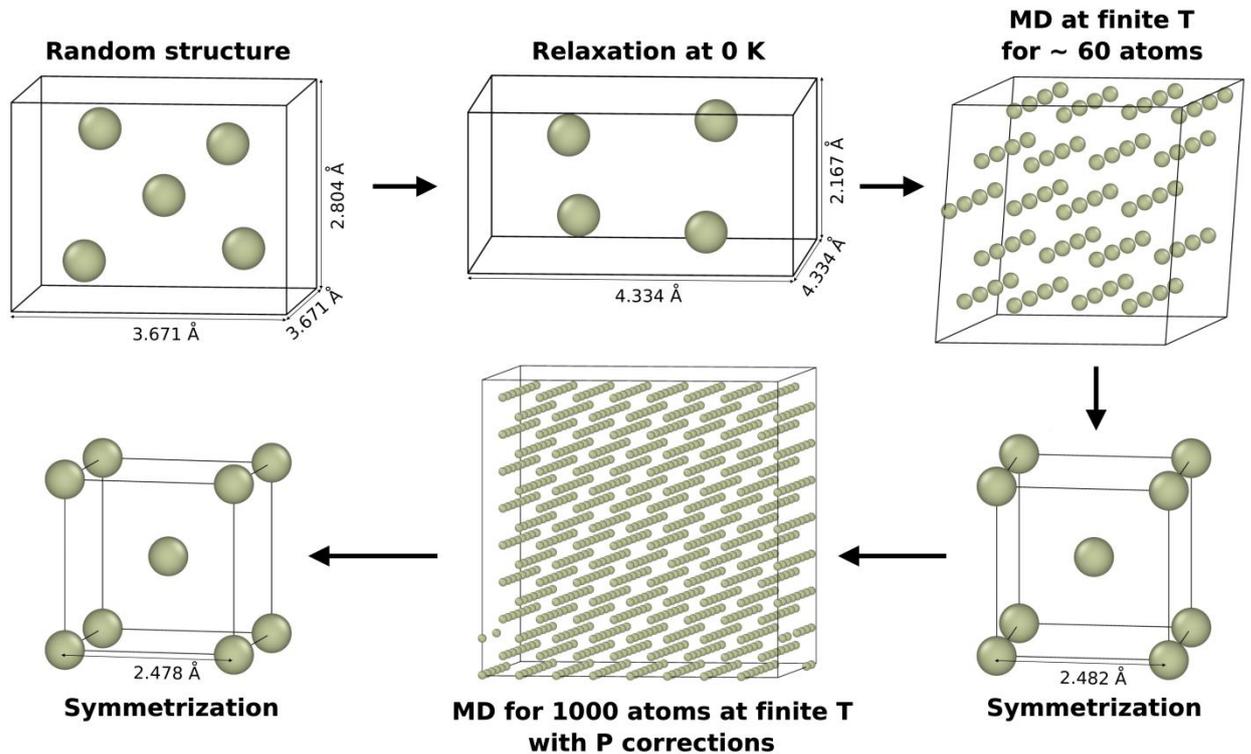

Figure 2. Evolution examples of randomly generated Al structure after relaxation at finite temperature within T-USPEX.

The model potential has two kinds of errors (relative to *ab initio* calculations) – a small error in density (which we compensate by a shift of pressure to bring pressure in exact agreement with *ab initio* result) and (at a given density) a small error in the free energy (which we correct for by thermodynamic perturbation theory). The pressure correction turned out to be several GPa at most (Fig. S3); at the simulated pressure of 400 GPa it varied from 5 to 12 GPa. In the calculation at 0 GPa and 300 K 15 structures (out of 20) transformed to *fcc*-Al with free energies equal to (-3.8474 ± 0.0002) eV/atom and densities (2.660 ± 0.0044) g/cm$^3$ , 1 structure to *I4/mmm* phase with free energy of -3.7490 eV/atom (clearly, this structure is not competitive), the other 4 structures being defective supercells (Table S1). We find that in the P-T region of interest (temperature up to 5000 K and pressures up to 400 GPa) only fcc, hcp and bcc structures have stability fields.

In order to calculate the P-T phase diagram, we used the *fcc*, *hcp* and *bcc* structures as seeds and performed T-USPEX calculations close to the previously calculated phase transition boundaries[40,42]. These calculations can be viewed as computer experiments akin to real experiments sampling the phase diagram. The resulting P-T phase diagram for Al is presented in Fig. 3 (see also Supplementary Fig. S4). Clearly, the *hcp* phase occupies a much larger region of the phase diagram than in previous studies[40–42]. Our fully anharmonic result should be more reliable than previous calculations[40–42] based on the quasiharmonic approximation.

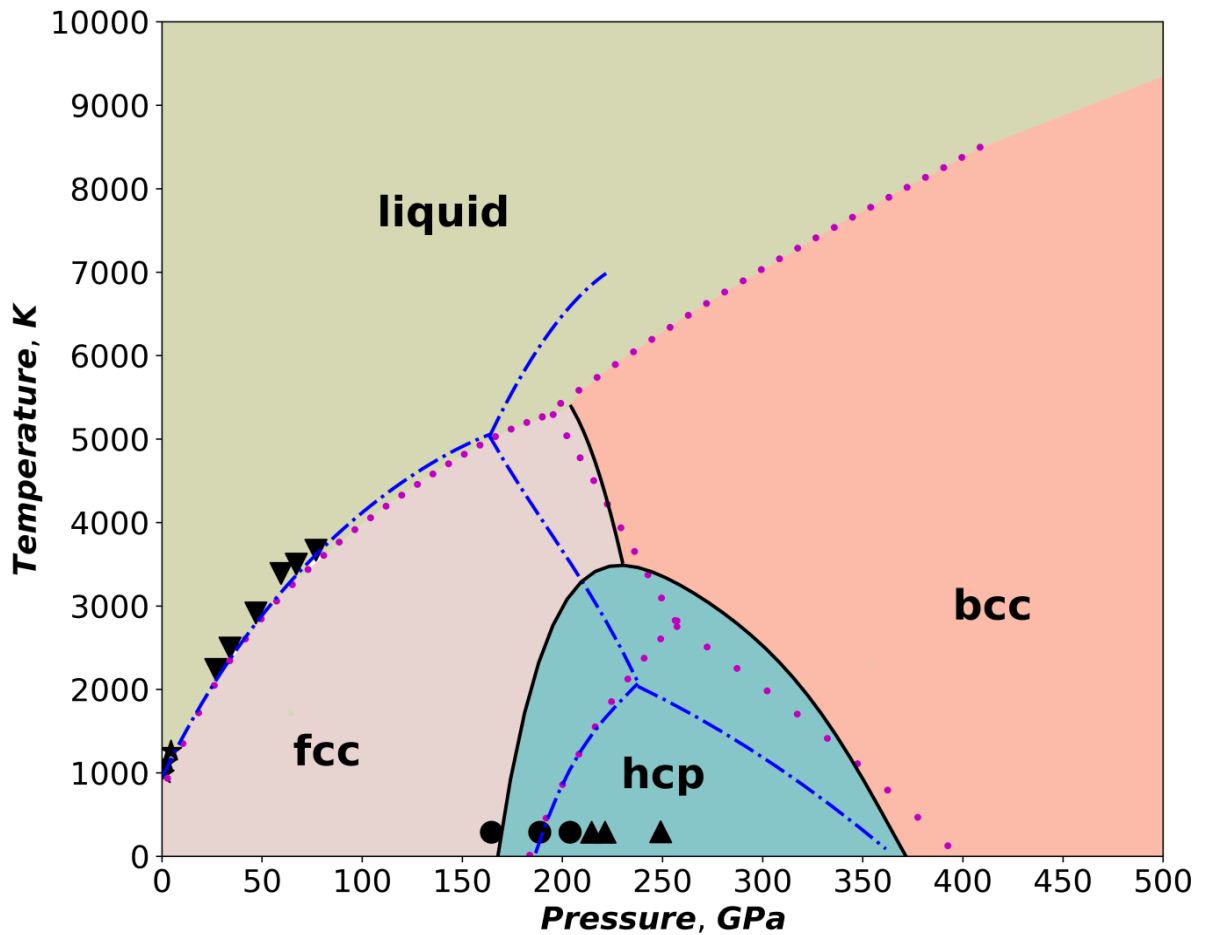

Figure 3. P-T phase diagram of Al. Black solid lines correspond to phase boundaries calculated in this work with T-USPEX. Black stars, triangles and circles – experimental data from Ref. [44,45]. Blue line is from Ref. [40], orange dotted line – from Ref. [42].

Until now, an open question is which phase of iron (bcc or hcp or perhaps some new structure) is stable at conditions of the Earth's core (pressures in the range 330-350 GPa and temperatures 5400-5700 K). It is difficult to conduct experiments at such extreme conditions, and available experiments give different answers[11–14]. In order to address this problem, we ran T-USPEX to find the most stable phase of iron at the pressure of 330 GPa and temperature of 5700 K. In the first generation consisting of 20 random structures, 7 structures relaxed into the *hcp* phase with the Gibbs free energy 6.080 ± 0.006 eV/atom, 1 – into the double *hcp* structure with the free energy of 6.094 eV/atom, 1 – into the *bcc* structure with free energy of 6.132 eV/atom. Thus, *hcp* structure is stable for iron at P-T conditions corresponding to the Earth's inner core, in agreement with results from Ref. 13.

Using T-USPEX, we explored the main Earth-forming compound – $MgSiO_3$ (comprising ~40 vol.% of our entire planet, or ~80 vol.% of its lower mantle) – searching for phases stable at conditions of the lowermost mantle. Throughout most of the lower mantle this compound exists as bridgmanite, a perovskite-type modification, while in the lowermost D" layer a post-perovskite polymorph is stable[15,16]. Here we followed a scheme somewhat different from what we used above, to illustrate the flexibility of our approach. For crystal structure relaxation and free energy calculations we used the classical interatomic potential developed in Ref. [46], which showed good agreement with experimental crystal structures and many properties. Here, using this interatomic potential, we ran T-USPEX for 20 generations (each containing 20 structures) at

135 GPa and 2700 K. Instead of applying expensive pressure and free energy corrections within crystal structure prediction runs, we ran highly efficient T-USPEX searches based solely on this interatomic potential, then selected structures with the lowest free energies and for these performed very accurate crystal structure and free energies calculations as described above.

In the T-USPEX search at 135 GPa and 4000 K, already in the first generation we saw post-perovskite (pPv) *Cmcm*-$MgSiO_3$ as the most stable structure, and in the second generation – bridgmanite, the phase with *Pbnm*-perovskite (Pv) structure, and no other competitive structures with up to 80 atoms in the primitive cell. Then we obtained accurate structures and their free energies using pressure and free energy corrections as described above; this was done at 120 GPa and 1500 K, 2000 K and 2500 K; at 130 GPa and 500 K, 1500 K, 3000 K, 4000 K. At 120 GPa pPv phase was stable up to 2000 K, at 130 GPa – up to 3700 K. Using these values the phase transition boundary between Pv and pPv $MgSiO_3$ phases was calculated (Fig. 4). Our calculated Pv-pPv phase equilibrium boundary is in good agreement with the results of previous theoretical[16,17] and experimental[15,18,19,47] works (see Fig. 4) and has a positive Clapeyron slope of +5.88 MPa/K. This value is important for calculating the heat flow from the core into the mantle and for determining the thermal structure of the lowermost mantle[19]. Here we give its most accurate theoretical calculation, for the first time taking into account the anharmonic part of the free energy.

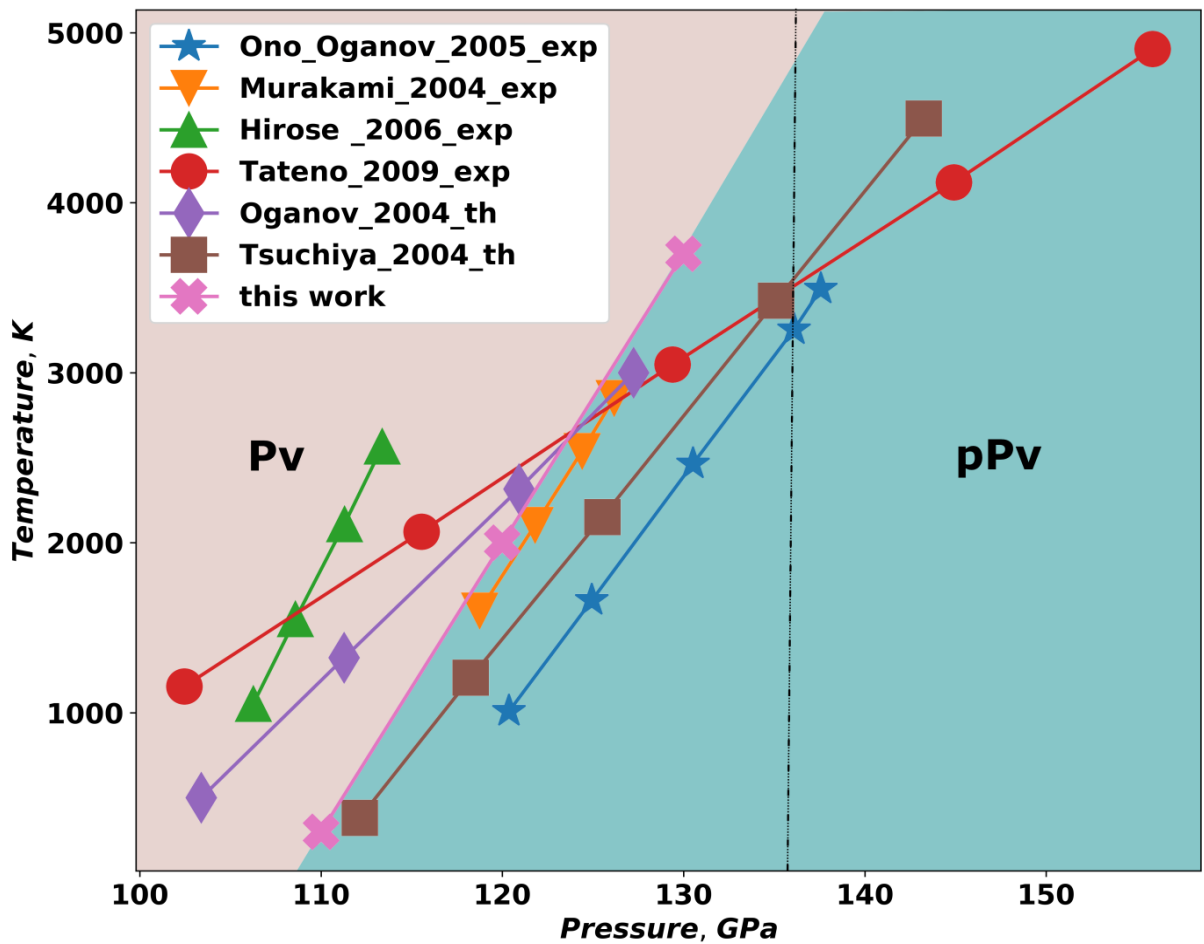

Figure 4. P-T phase diagram of $MgSiO_3$. Horizontal dashed line shows the pressure of the Earth's core-mantle boundary.

Next, we applied our method to WB – a refractory ultrahard compound, possessing a temperature-induced phase transition and characterized by a non-trivial chemical bonding and with a lack of good classical force field. The temperature-induced phase transition from the low-temperature α-phase into the high-temperature β-phase takes place at approximately 2100-2700

K[48–50]. We ran T-USPEX at 0 GPa and 2000 K with MTP force field, and we obtained both $I4_1/amd$-WB (α-phase) and *Cmcm*-WB (β-phase) and a hitherto unknown low-energy *Pnma* structure of WB to have the lowest free energies (within a few meV/atom of each other, see Supplementary Table S2). In order to find stability regions of the discovered tungsten monoborides, their equilibrium crystal structures and free energies were calculated with higher accuracy in the temperature range of 300-3000 K at zero pressure. Our calculations show that the α-β transition takes place at 2789 K, in agreement with experimental results[49], while *Pnma*-WB remains a low-energy metastable phase at all temperatures at zero pressure.

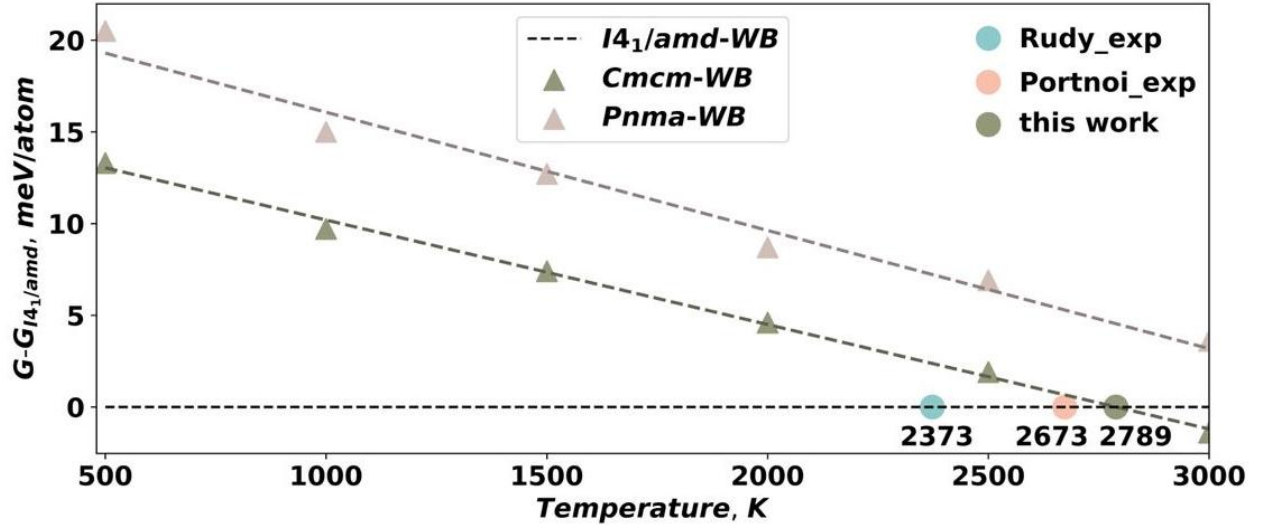

Fig. 5. Gibbs free energies of *Cmcm* and *Pnma* structures of WB (relative to $I4_1/amd$).

Until now, finite-temperature crystal structure prediction was thought to be prohibitively expensive. Here we showed a computationally efficient approach solving this problem, and illustrated its power by applying it to four very different substances: simple metal (Al at high pressures and temperatures), a transition metal (Fe in Earth's core), a geologically important silicate ($MgSiO_3$ at conditions of the lowermost mantle), and a hard refractory compound WB with a temperature-induced phase transition. We show how this method can be used, in the spirit of a computational experiment akin to a real experiment, to probe phase diagrams, resolve long-standing disputes, and open new horizons in the search for unique materials with improved properties. Our method takes advantage of increasing simplifications of the free energy landscape on increasing temperature, and of reliable machine learning force fields and thermodynamic perturbation theory which allow highly accurate and affordable calculations of free energies. This development enables crystal structure prediction at any temperature for ordered materials. Further work is needed to extend this approach to configurationally or magnetically disordered systems.

**Acknowledgements**

I.A.K. thanks Russian Science Foundation (grant No. 19-73-00237) for financial support. Work of A.R.O. is supported by Russian Science Foundation (grant 19-72-30043).

**References**

1. Laio, A. & Parrinello, M. Escaping free-energy minima. *Proc. Natl. Acad. Sci. U. S. A.* **99**, 12562–12566 (2002).
2. Pickard, C. J. & Needs, R. J. Ab initio random structure searching. *Journal of Physics Condensed Matter* **23**, 053201 (2011).
3. Glass, C. W., Oganov, A. R. & Hansen, N. USPEX—Evolutionary crystal structure prediction. *Comput. Phys. Commun.* **175**, 713–720 (2006).

4. Oganov, A. R., Lyakhov, A. O. & Valle, M. How Evolutionary Crystal Structure Prediction Works—and Why. *Acc. Chem. Res.* **44**, 227–237 (2011).
5. Lyakhov, A. O., Oganov, A. R., Stokes, H. T. & Zhu, Q. New developments in evolutionary structure prediction algorithm USPEX. *Comput. Phys. Commun.* **184**, 1172–1182 (2013).
6. Goedecker, S. Minima hopping: An efficient search method for the global minimum of the potential energy surface of complex molecular systems. **120**, 9911–9917 (2004).
7. Wang, Y., Lv, J., Zhu, L. & Ma, Y. CALYPSO: A method for crystal structure prediction. *Comput. Phys. Commun.* **183**, 2063–2070 (2012).
8. Lonie, D. C. & Zurek, E. XtalOpt: An open-source evolutionary algorithm for crystal structure prediction. *Comput. Phys. Commun.* **182**, 372–387 (2011).
9. Oganov, A. R., Pickard, C. J., Zhu, Q. & Needs, R. J. Structure prediction drives materials discovery. *Nature Reviews Materials* **4**, 331–348 (2019).
10. Allahyari, Z. & Oganov, A. R. Coevolutionary search for optimal materials in the space of all possible compounds. *npj Comput. Mater.* **6**, 1–10 (2020).
11. Saxena, S. K. *et al.* Synchrotron x-ray study of iron at high pressure and temperature. *Science (80-. ).* **269**, 1703–1704 (1995).
12. Belonoshko, A. B., Ahuja, R. & Johansson, B. Stability of the body-centred-cubic phase of iron in the Earth's inner core. *Nature* **424**, 1032–1034 (2003).
13. Tateno, S., Hirose, K., Ohishi, Y. & Tatsumi, Y. The structure of iron in earth's inner core. *Science (80-. ).* **330**, 359–361 (2010).
14. Vočadlo, L. *et al.* Possible thermal and chemical stabilization of body-centered-cubic iron in the Earth's core. *Nature* **424**, 536–539 (2003).
15. Murakami, M., Hirose, K., Kawamura, K., Sata, N. & Ohishi, Y. Post-Perovskite Phase Transition in MgSiO3. *Science (80-. ).* **304**, 855–858 (2004).
16. Oganov, A. R. & Ono, S. Theoretical and experimental evidence for a post-perovskite phase of MgSiO3 in Earth's D″ layer. *Nature* **430**, 445–448 (2004).
17. Tsuchiya, T., Tsuchiya, J., Umemoto, K. & Wentzcovitch, R. M. Phase transition in MgSiO3 perovskite in the earth's lower mantle. *Earth Planet. Sci. Lett.* **224**, 241–248 (2004).
18. Ono, S. & Oganov, A. R. In situ observations of phase transition between perovskite and CaIrO3-type phase in MgSiO3 and pyrolitic mantle composition. *Earth Planet. Sci. Lett.* **236**, 914–932 (2005).
19. Tateno, S., Hirose, K., Sata, N. & Ohishi, Y. Determination of post-perovskite phase transition boundary up to 4400 K and implications for thermal structure in D″ layer. *Earth Planet. Sci. Lett.* **277**, 130–136 (2009).
20. Oganov, A. R. Crystal structure prediction: reflections on present status and challenges. *Faraday Discuss.* **211**, 643–660 (2018).
21. Behler, J. & Parrinello, M. Generalized neural-network representation of high-dimensional potential-energy surfaces. *Phys. Rev. Lett.* **98**, 146401 (2007).
22. Bartók, A. P., Payne, M. C., Kondor, R. & Csányi, G. Gaussian Approximation Potentials: The Accuracy of Quantum Mechanics, without the Electrons. *Phys. Rev. Lett.* **104**, 136403 (2010).
23. Shapeev, A. V. Moment Tensor Potentials: A Class of Systematically Improvable Interatomic Potentials. *Multiscale Model. Simul.* **14**, 1153–1173 (2016).
24. Li, Z., Kermode, J. R. & De Vita, A. Molecular Dynamics with On-the-Fly Machine Learning of Quantum-Mechanical Forces. *Phys. Rev. Lett.* **114**, 096405 (2015).
25. Kruglov, I., Sergeev, O., Yanilkin, A. & Oganov, A. R. Energy-free machine learning force field for aluminum. *Sci. Rep.* **7**, 8512 (2017).
26. Podryabinkin, E. V., Tikhonov, E. V., Shapeev, A. V. & Oganov, A. R. Accelerating crystal structure prediction by machine-learning interatomic potentials with active learning. *Phys. Rev. B* **99**, (2019).

27. Podryabinkin, E. V. & Shapeev, A. V. Active learning of linearly parametrized interatomic potentials. *Comput. Mater. Sci.* **140**, 171–180 (2017).
28. Artrith, N., Morawietz, T. & Behler, J. High-dimensional neural-network potentials for multicomponent systems: Applications to zinc oxide. *Phys. Rev. B* **83**, 153101 (2011).
29. Behler, J. Representing potential energy surfaces by high-dimensional neural network potentials. *J. Phys. Condens. Matter* **26**, 183001 (2014).
30. Szlachta, W. J., Bartók, A. P. & Csányi, G. Accuracy and transferability of Gaussian approximation potential models for tungsten. *Phys. Rev. B* **90**, 104108 (2014).
31. Dolgirev, P. E., Kruglov, I. A. & Oganov, A. R. Machine learning scheme for fast extraction of chemically interpretable interatomic potentials. *AIP Adv.* **6**, 085318 (2016).
32. Kruglov, I. A., Yanilkin, A., Oganov, A. R. & Korotaev, P. Phase diagram of uranium from ab initio calculations and machine learning. *Phys. Rev. B* **100**, 174104 (2019).
33. Zuo, Y. *et al.* A Performance and Cost Assessment of Machine Learning Interatomic Potentials. *J. Phys. Chem. A* (2019). doi:10.1021/acs.jpca.9b08723
34. Korotaev, P., Novoselov, I., Yanilkin, A. & Shapeev, A. Accessing thermal conductivity of complex compounds by machine learning interatomic potentials. *Phys. Rev. B* **100**, 144308 (2019).
35. Ladygin, V. V., Korotaev, P. Y., Yanilkin, A. V. & Shapeev, A. V. Lattice dynamics simulation using machine learning interatomic potentials. *Comput. Mater. Sci.* **172**, 109333 (2020).
36. Novikov, I. S., Gubaev, K., Podryabinkin, E. V. & Shapeev, A. V. The MLIP package: Moment Tensor Potentials with MPI and Active Learning. (2020).
37. Straatsma, T. P. & Berendsen, H. J. C. Free energy of ionic hydration: Analysis of a thermodynamic integration technique to evaluate free energy differences by molecular dynamics simulations. *J. Chem. Phys.* **89**, 5876–5886 (1988).
38. Freitas, R., Asta, M. & de Koning, M. Nonequilibrium free-energy calculation of solids using LAMMPS. *Comput. Mater. Sci.* **112**, 333–341 (2016).
39. Landau, L. D. & Lifshitz, E. M. Statistical Physics, Part 1, Vol. 5. *Course Theor. Phys.* **3**, (1994).
40. Sjostrom, T., Crockett, S. & Rudin, S. Multiphase aluminum equations of state via density functional theory. *Phys. Rev. B* **94**, 144101 (2016).
41. Tambe, M. J., Bonini, N. & Marzari, N. Bulk aluminum at high pressure: A first-principles study. *Phys. Rev. B - Condens. Matter Mater. Phys.* **77**, 172102 (2008).
42. Kudasov, Y. B. *et al.* Lattice dynamics and phase diagram of aluminum at high temperatures. *J. Exp. Theor. Phys.* **117**, 664–671 (2013).
43. Akahama, Y., Nishimura, M., Kinoshita, K., Kawamura, H. & Ohishi, Y. Evidence of a fcc-hcp transition in aluminum at multimegabar pressure. *Phys. Rev. Lett.* **96**, 045505 (2006).
44. Grigoriev, I. & Meilikhov, E. *Handbook of Physical Quantities (November 25, 1996 ed.).* (Energoatomizdat, Moscow, 1991; CRC Press, Boca Raton, Florida, United States, 1996).
45. Boehler, R. & Ross, M. Melting curve of aluminum in a diamond cell to 0.8 Mbar: Implications for iron. *Earth Planet. Sci. Lett.* **153**, 223–227 (1997).
46. Oganov, A. R., Brodholt, J. P. & Price, G. D. Comparative study of quasiharmonic lattice dynamics, molecular dynamics and Debye model applied to MgSiO3 perovskite. *Phys. Earth Planet. Inter.* **122**, 277–288 (2000).
47. Hirose, K., Sinmyo, R., Sata, N. & Ohishi, Y. Determination of post-perovskite phase transition boundary in MgSiO$_3$ using Au and MgO pressure standards. *Geophys. Res. Lett.* **33**, n/a-n/a (2006).
48. Rudy, E. *Experimental Phase Equilibria of Selected Binary, Ternary, and Higher Order Systems. Part 5. the Phase Diagram W-B-C.* (1970).
49. Portnoi, K. I. & Romashov, V. M. Phase diagram of the system rhenium-boron. *Sov. Powder Metall. Met. Ceram.* **7**, 112–114 (1968).

50. Kvashnin, A. G. & Samtsevich, A. I. Phase Transitions in Tungsten Monoborides. *JETP Lett.* **111**, 343–349 (2020).
51. Perdew, J. P., Burke, K. & Ernzerhof, M. Generalized gradient approximation made simple. *Phys. Rev. Lett.* **77**, 3865–3868 (1996).
52. Kresse, G. & Hafner, J. *Ab initio* molecular dynamics for liquid metals. *Phys. Rev. B* **47**, 558–561 (1993).
53. Kresse, G. & Hafner, J. *Ab initio* molecular-dynamics simulation of the liquid-metal–amorphous-semiconductor transition in germanium. *Phys. Rev. B* **49**, 14251–14269 (1994).
54. Kresse, G. & Furthmüller, J. Efficient iterative schemes for *ab initio* total-energy calculations using a plane-wave basis set. *Phys. Rev. B* **54**, 11169–11186 (1996).
55. Plimpton, S. Fast Parallel Algorithms for Short-Range Molecular Dynamics. *J. Comput. Phys.* **117**, 1–19 (1995).
56. Bushlanov, P. V., Blatov, V. A. & Oganov, A. R. Topology-based crystal structure generator. *Comput. Phys. Commun.* **236**, 1–7 (2019).
57. Korotaev, P., Belov, M. & Yanilkin, A. Reproducibility of vibrational free energy by different methods. *Comput. Mater. Sci.* **150**, 47–53 (2018).
58. Hjorth Larsen, A. *et al.* The atomic simulation environment - A Python library for working with atoms. *Journal of Physics Condensed Matter* **29**, (2017).
59. Togo, A. & Tanaka, I. Spglib: a software library for crystal symmetry search. (2018).

**Methods**

Crystal structure is performed here using a finite-temperature extension of the USPEX method/code[3–5] at the GGA (generalized gradient approximation)[51] level of density functional theory (DFT). DFT calculations are done using the VASP code[52–54] with high level of convergence (see Supplementary Materials for details). Note that for metals the electronic free energy was calculated and included in the total free energy. Molecular dynamics with classical or ML force fields was done using LAMMPS code[55].

The initial population of structures was created using symmetric[5] and topological[56] random structure generators. Each produced structure was then relaxed using molecular dynamics in the NPT ensemble. Free energy is calculated using thermodynamic integration as the most accurate method[57]. To perform structure relaxations and free energy calculations with molecular dynamics, large supercells are used (more than 1000 atoms), which can only be done using a force field. If for a given system a good classical force field is known, then it can be applied in molecular dynamics simulations (as we did for $MgSiO_3$). However, for most systems there is no classical force field which can work in a vast configurational space (created by random structures from T-USPEX). In this case we use a machine learning force field trained on *ab initio* data. To achieve ultimate accuracy, we separately train a ML force field for each structure at a given P-T-point, deliberately sacrificing efficiency for the sake of accuracy. We used MTP machine learning force field[23], due to its outstanding performance[33] and ability to perform active learning[27], i.e. re-train the force field whenever a structure different from its original training set is encountered.

Since MTP is trained on DFT data, first, a newly generated crystal structure is relaxed using DFT at 0 K, after which it is relaxed at finite temperature using NPT molecular dynamics and ML force field (which is trained on the NPT ab initio molecular dynamics trajectory in an approximately isotropic "small" supercell with ~60 atoms). Even the best ML force fields give rise to non-negligible errors in pressure estimations, and it will affect densities of the final structures and the PV-term in the free energy. For this reason we calculate the pressure correction: we perform a molecular dynamics simulation in the NVE ensemble using the force field for 20 ps, and intermediate structures are saved every 1 ps – for each of which we calculate the pressure using DFT, and pressure correction is taken as ensemble average.

In the next step the previously obtained structure is replicated to form a supercell with ~1000 atoms ("large" supercell), which is relaxed using NPT molecular dynamics for 40 ps, taking pressure correction into account. Performing ensemble averaging of supercell vectors and fractional coordinates of each atom, we then obtain the crystal structure, the symmetry and minimal unit cell of which is then determined and communicated to T-USPEX using ASE[58] and Spglib[59] libraries.. Structures with large and defective unit cells are discarded. For the remaining structures, we compute Gibbs free energies as described in the main text.

# Supplementary materials

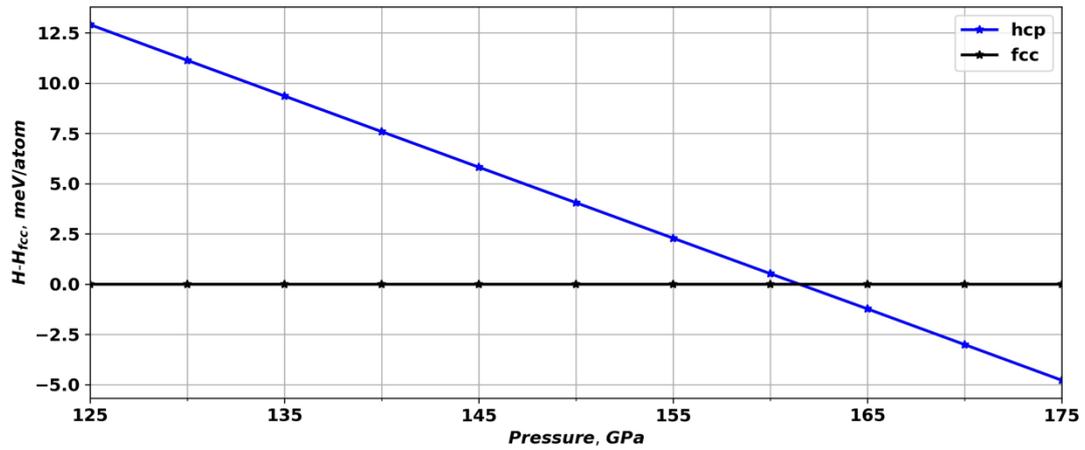

Figure S1. Enthalpy difference between *hcp*- and *fcc*-Al phases as a function of pressure.

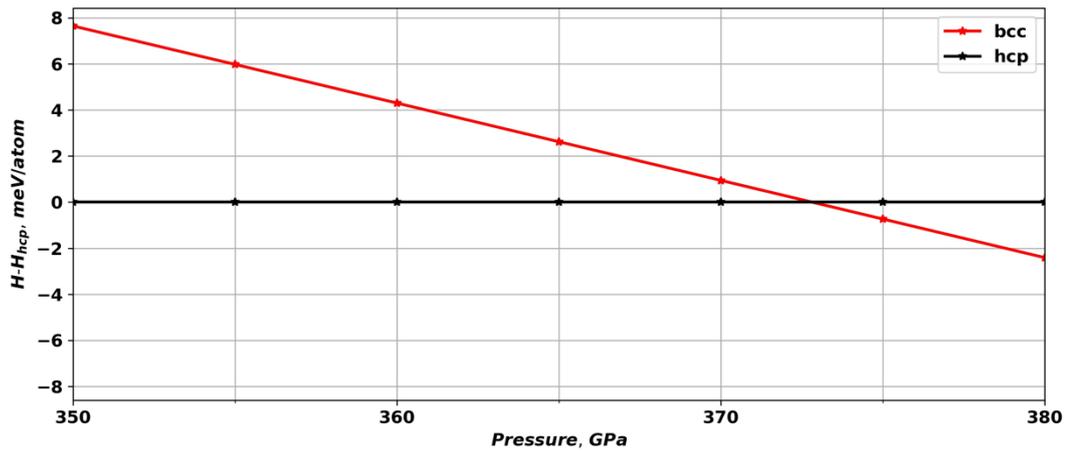

Figure S2. Enthalpy difference between *bcc*- and *hcp*-Al phases as a function of pressure.

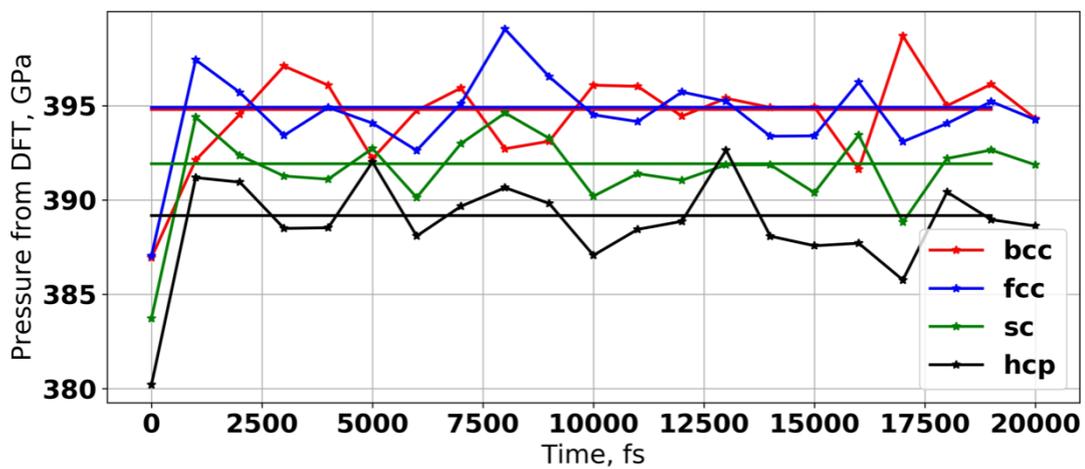

Figure S3. Pressures for intermediate structures from NVE molecular dynamics run with MTP potentials for different Al structures.

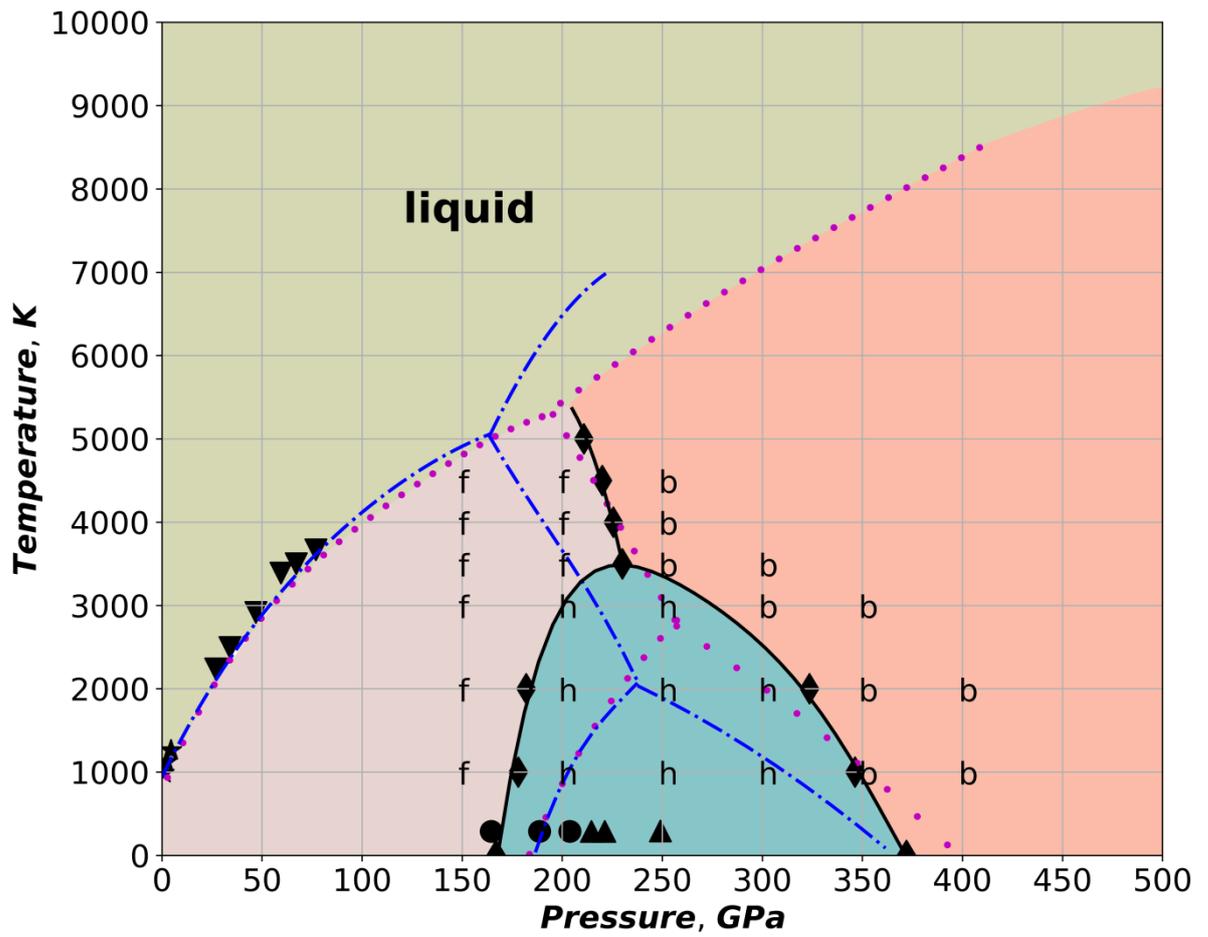

Figure S4. P-T phase diagram of Al. Black solid lines correspond to phase transition boundaries calculated in this work with T-USPEX. Black stars, triangles and circles – experimental data from Ref. Blue line is from Ref. [37], orange dotted line – from Ref. [39]. Letters "f", "h" and "b" correspond to the most stable phases (from T-USPEX) at given P-T conditions. Black diamonds show calculated transition points between different phases.

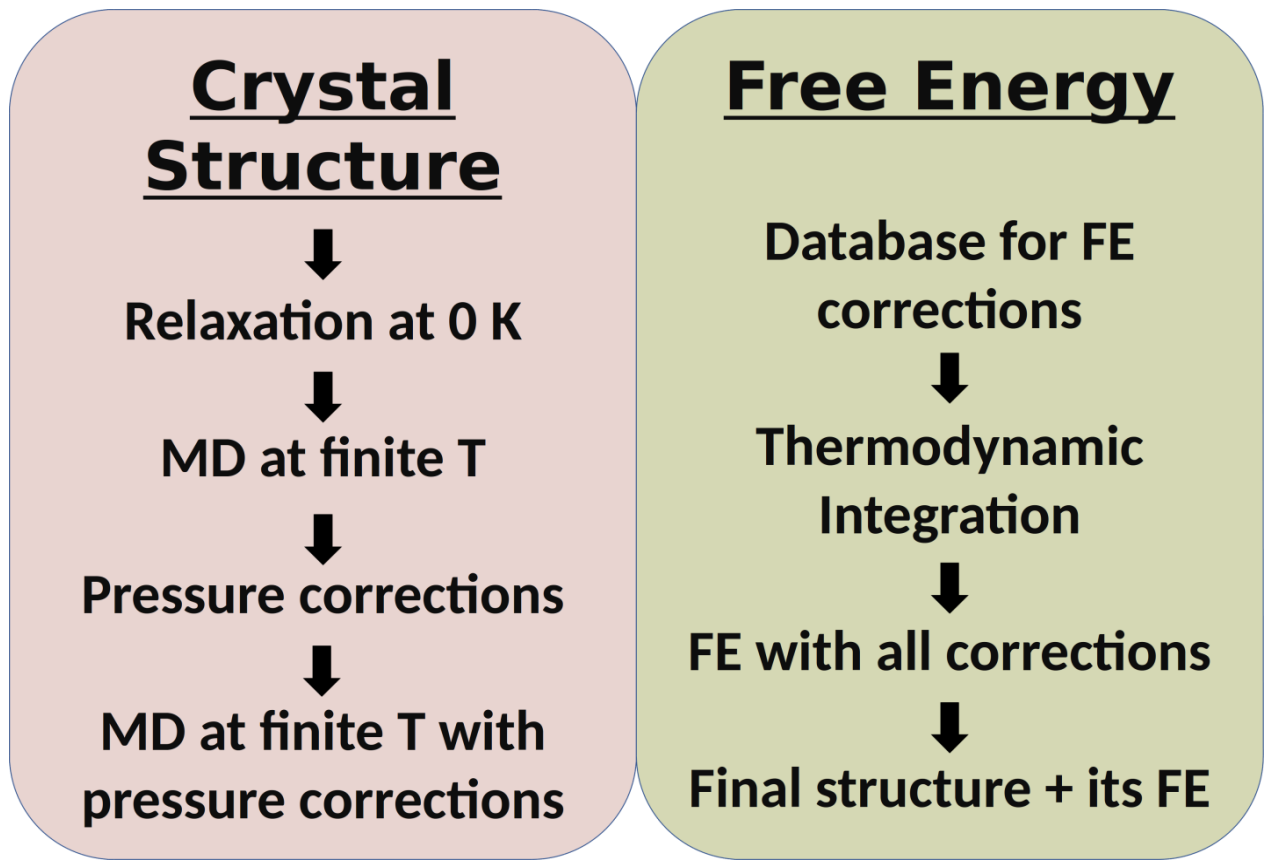

Figure S5. Flow chart of the T-USPEX method. MD = molecular dynamics, FE = free energy.

Table S1. Total Gibbs free energies (G) with all contributions ($F_{Helm}$ - Helmholtz free energy from thermodynamic integration (to Einstein crystal), $FE_{corr1}$ and $FE_{corr2}$ - 1st and 2nd order corrections from thermodynamic perturbation theory, $E_{Eins}$ – energy of Einstein crystal) for Al structures from T-USPEX at 0 GPa and 300 K. All energies are given in eV/atom.

| Space group | $N_{atoms}$ | $F_{Helm}$ | $FE_{corr1}$ | $FE_{corr2}$ | pV | $E_{Eins}$ | G |
|---|---|---|---|---|---|---|---|
| $Fm\bar{3}m$ | 4 | -3.8327 | 0.0025 | 0.0000 | 0.0000 | -0.0171 | -3.8473 |
| $Fm\bar{3}m$ | 4 | -3.8159 | -0.0144 | 0.0000 | 0.0000 | -0.0171 | -3.8474 |
| $Fm\bar{3}m$ | 4 | -3.8390 | 0.0086 | 0.0000 | 0.0000 | -0.0171 | -3.8475 |
| $Fm\bar{3}m$ | 4 | -3.8090 | -0.0211 | 0.0000 | 0.0000 | -0.0171 | -3.8473 |
| $Fm\bar{3}m$ | 4 | -3.8095 | -0.0205 | 0.0000 | 0.0000 | -0.0171 | -3.8472 |
| $Fm\bar{3}m$ | 4 | -3.8286 | -0.0014 | 0.0000 | 0.0000 | -0.0171 | -3.8471 |
| $Fm\bar{3}m$ | 4 | -3.8237 | -0.0066 | 0.0000 | 0.0000 | -0.0171 | -3.8474 |
| $Fm\bar{3}m$ | 4 | -3.8224 | -0.0080 | 0.0000 | 0.0000 | -0.0171 | -3.8476 |
| $Fm\bar{3}m$ | 4 | -3.8060 | -0.0244 | 0.0000 | 0.0000 | -0.0171 | -3.8475 |
| $Fm\bar{3}m$ | 4 | -3.8279 | -0.0025 | 0.0000 | 0.0000 | -0.0171 | -3.8475 |
| $Fm\bar{3}m$ | 4 | -3.8091 | -0.0209 | 0.0000 | 0.0000 | -0.0171 | -3.8472 |
| $Fm\bar{3}m$ | 4 | -3.8199 | -0.0105 | 0.0000 | 0.0000 | -0.0171 | -3.8475 |
| $Fm\bar{3}m$ | 4 | -3.8101 | -0.0203 | 0.0000 | 0.0000 | -0.0171 | -3.8475 |
| $Fm\bar{3}m$ | 4 | -3.8307 | 0.0005 | 0.0000 | 0.0000 | -0.0171 | -3.8474 |
| $I4/mmm$ | 14 | -3.7393 | 0.0075 | 0.0000 | 0.0000 | -0.0171 | -3.7490 |
| $Fm\bar{3}m$ | 4 | -3.8293 | -0.0010 | 0.0000 | 0.0000 | -0.0171 | -3.8475 |

Table S2. Total Gibbs free energies (G) with all contributions ($F_{Helm}$ - Helmholtz free energy from thermodynamic integration (to Einstein crystal), $FE_{corr1}$ and $FE_{corr2}$ - 1$^{st}$ and 2$^{nd}$ order corrections from thermodynamic perturbation theory, $E_{Eins}$ – energy of Einstein crystal) for WB structures from T-USPEX at 0 GPa and 2000 K. All energies are given in eV/atom.

| Space group | $N_{atoms}$ | $F_{Helm}$ | $FE_{corr1}$ | $FE_{corr2}$ | pV | $E_{Eins}$ | G |
|---|---|---|---|---|---|---|---|
| $Fm\bar{3}m$ | 8 | -10.1914 | 0.0508 | 0.0000 | 0.0000 | -0.6136 | -10.7543 |
| $Cmcm$ | 8 | -10.3158 | -0.0236 | 0.0001 | 0.0000 | -0.6136 | -10.9531 |
| $Cmcm$ | 8 | -10.3208 | -0.0272 | 0.0001 | 0.0000 | -0.6136 | -10.9617 |
| $Cmcm$ | 8 | -10.3503 | 0.0011 | 0.0000 | 0.0000 | -0.6136 | -10.9629 |
| $I4_1/amd$ | 16 | -10.3339 | -0.0162 | 0.0000 | 0.0000 | -0.6136 | -10.9638 |
| $Pnma$ | 8 | -10.3307 | -0.0117 | 0.0001 | 0.0000 | -0.6136 | -10.9561 |